\pdfoutput=1

\documentclass[conference]{IEEEtran}
\usepackage{amsmath,amssymb,amsthm,epsfig,color,empheq,graphicx,graphics,balance}
\usepackage{enumerate,url,wasysym,epstopdf,enumitem,array}
\usepackage{xcolor}
\usepackage{amsfonts}
\usepackage{adjustbox}
\usepackage{multirow}
\usepackage{cite}
\usepackage{accents}

\usepackage{algorithm2e} 
\usepackage{float} 
\usepackage{bbold}

\DeclareMathOperator{\diag}{dg}

\newcommand \bzero{\mathbf{0}}
\newcommand \bone{\mathbf{1}}

\newcommand \bc{\mathbf{c}}

\newcommand \bq{\mathbf{q}}

\newcommand \bv{\mathbf{v}}

\newcommand \bz{\mathbf{z}}

\newcommand \bR{\mathbf{R}}

\newcommand \bX{\mathbf{X}}

\newcommand \balpha{\boldsymbol{\alpha}}

\newcommand \bdelta{\boldsymbol{\delta}}

\newcommand \btheta{\boldsymbol{\theta}}

\newcommand \blambda{\boldsymbol{\lambda}}

\newcommand \bsigma{\boldsymbol{\sigma}}

\newcommand \mcN{\mathcal{N}}

\newcommand \mcZ{\mathcal{Z}}

\newcommand \tbp{\tilde{\mathbf{p}}}
\newcommand \tbq{\tilde{\mathbf{q}}}

\newcommand \tbv{\tilde{\mathbf{v}}}

\newcommand \cbz{\check{\mathbf{z}}}

\newcommand \hbq{\hat{\mathbf{q}}}

\newcommand \hbz{\hat{\mathbf{z}}}

\newcommand \bbv{\bar{\mathbf{v}}}

\renewcommand{\d}[1]{\ensuremath{\operatorname{d}\!{#1}}}

\begin{document}

\title{A Chance-Constrained Optimal Design of Volt/VAR\\
Control Rules for Distributed Energy Resources}

\author{
\IEEEauthorblockN{Jinlei Wei}
\IEEEauthorblockA{Bradley Dept. of ECE}
\textit{Virginia Tech}\\
Blacksburg, VA\\
\texttt{jinlei@vt.edu}
\and
\IEEEauthorblockN{Sarthak Gupta}
\IEEEauthorblockA{Bradley Dept. of ECE}
\textit{Virginia Tech}\\
Blacksburg, VA\\
\texttt{gsarthak@vt.edu}
\and
\IEEEauthorblockN{Dionysios C. Aliprantis}
\IEEEauthorblockA{Elmore Family School of ECE}
\textit{Purdue University}\\
West Lafayette, IN\\
\texttt{dionysios@purdue.edu}
\and
\IEEEauthorblockN{Vassilis Kekatos}
\IEEEauthorblockA{Bradley Dept. of ECE}
\textit{Virginia Tech}\\
Blacksburg, VA\\
\texttt{kekatos@vt.edu}
}

\maketitle


\begin{abstract}
Deciding setpoints for distributed energy resources (DERs) via local control rules rather than centralized optimization offers significant autonomy. The IEEE Standard 1547 recommends deciding DER setpoints using Volt/VAR rules. Although such rules are specified as non-increasing piecewise-affine, their exact shape is left for the utility operators to decide and possibly customize per bus and grid conditions. To address this need, this work optimally designs Volt/VAR rules to minimize ohmic losses on lines while maintaining voltages within allowable limits. This is practically relevant as excessive reactive injections could reduce equipment's lifetime due to overloading. We consider a linearized single-phase grid model. Even under this setting, optimal rule design (ORD) is technically challenging as Volt/VAR rules entail mixed-integer models, stability implications, and uncertainties in grid loading. Uncertainty is handled by minimizing the average losses under voltage chance constraints. To cope with the piecewise-affine shape of the rules, we build upon our previous reformulation of ORD as a deep learning task. A recursive neural network (RNN) surrogates Volt/VAR dynamics and thanks to back-propagation, we expedite this chance-constrained ORD. RNN weights coincide with rule parameters, and are trained using primal-dual decomposition. Numerical tests corroborate the efficacy of this novel ORD formulation and solution methodology.
\end{abstract}


\section{Introduction}
\allowdisplaybreaks
\footnote{Work supported in part by US NSF grant 2034137.}DERs are vital for integrating renewables as well as enhancing grid efficiency and reliability. To achieve these goals however, operators need to carefully design reactive power control by DERs for regulating voltages while reducing ohmic losses. For DERs to operate autonomously, the IEEE Std. 1547 provisions Volt/VAR rules according to which DERs inject reactive power based on local voltage~\cite{IEEE1547}. However, the standard does not specify the exact shape of these rules. In response to this need, this work develops a chance-constrained solution to the task of \emph{optimal rule design} (ORD).

Albeit simple to implement, local rules are known to be subpar compared to setpoints computed centrally by the optimal power flow (OPF)~\cite{9091863}. This necessitates customizing the rules possibly on a per-bus basis to achieve feeder-wide objectives. To increase autonomy, the operator should solve ORD relatively infrequently using grid loading scenarios anticipated for the next two hours or so. References~\cite{Jabr18,SKL19GM} have designed piecewise-linear Watt/VAR rules using mixed-integer linear programs (MILP). Designing Volt/VAR rules is more challenging as they give rise to nonlinear dynamics and call for mixed-integer nonlinear programs (MINLP)~\cite{GCK23}. The stability and equilibria of the dynamics induced by Volt/VAR rules have been studied in~\cite{Pedram13,9091863}, wherein rules were taken as fixed. Reference~\cite{9796576} incorporated fixed rules into the OPF. References~\cite{PaudyalMISOCP,Paudyal-LinDist3Flow} suggest ORD formulations of varying complexity including MILPs and mixed-integer second-order cone programs. Nonetheless, stability constraints and constraints required by IEEE Std. 1547 are not enforced, and rules are designed based on a single scenario. An adaptive strategy for updating Volt/VAR rules is put forth in~\cite{8365842}. Reference~\cite{9781808} designed affine Volt/VAR rules operating in conjuction with Volt/Watt rules. To deal with variability in grid conditions, work~\cite{Baker18} pursued a robust approach and designed stable yet affine Volt/VAR rules to minimize voltage deviations under a polytopic uncertainty set. The detailed shape of Volt/VAR rules is optimized via an MINLP and relaxation heuristics in~\cite{9609090}, yet constraints guaranteeing stability are not enforced. In a nutshell, prior works often ignore the deadband and saturation of Volt/VAR rules, and/or stability concerns to simplify ORD.


We have previously designed Volt/VAR rules to minimize squared voltage deviations from unity under stability constraints. In~\cite{MGCK23}, ORD was tackled as a bilevel program via projected gradient descent. The same ORD was later expedited through a genuine reformulation to a deep learning task in~\cite{GCK23}, which also extended analysis and synthesis of Volt/VAR rules to multiphase feeders. The idea was to design a RNN that accepts loading conditions as inputs, rule parameters as weights, and computes equilibrium voltages at its output. If the RNN is trained so its outputs approach unity for all scenarios, we have designed rules that regulate voltages indeed. Designing rules as NNs has been also suggested in~\cite{Baosen22,10106014}. The key difference is that our RNN is used only as a computational tool to speed up ORD and is never implemented on the field.

The contribution of this work is to design stable IEEE 1547 Volt/VAR rules to minimize losses under voltage constraints. This is practically pertinent as excessive reactive injections to regulate voltage could increase apparent power flows and reduce the equipment's lifetime. Regulating voltage and minimizing losses have been shown to be contradictory goals~\cite{Turitsyn11}. This work adopts common OPF renditions that decide DER setpoints to minimize losses while enforcing constraints on voltages. Designing Volt/VAR rules is technically non-trivial since ensuring stable rules gives rise to MINLPs. Similarly to~\cite{Ayyagari17,Li18}, uncertainty is accounted for using sample approximation and minimize averaged losses under chance constraints on voltages. Distinct from~\cite{Ayyagari17,Li18}, this work designs control rules rather than static setpoints. It extends the RNN approach of~\cite{GCK23} to deal with chance constraints. To enforce voltage constraints in probability and stability constraints at all times, we develop a primal-dual decomposition to train the RNN emulating Volt/VAR dynamics. Numerical tests on the IEEE 37-bus feeder showcase the proposed ORD scales favorably with the number of scenarios, and the obtained rules offer a tunable trade-off between losses and voltage regulation.

The remainder of this work is organized as follows. Section~\ref{sec:problem} formulates ORD after reviewing rule and grid models. Section~\ref{sec:method} presents and trains the RNN emulating Volt/VAR dynamics via primal-dual decomposition. Section~\ref{sec:tests} evaluates the proposed ORD. Conclusions are drawn in Section~\ref{sec:conclusions}.

\section{Problem Formulation}\label{sec:problem}


\begin{figure}[t]
	\centering
	\includegraphics[scale=0.3]{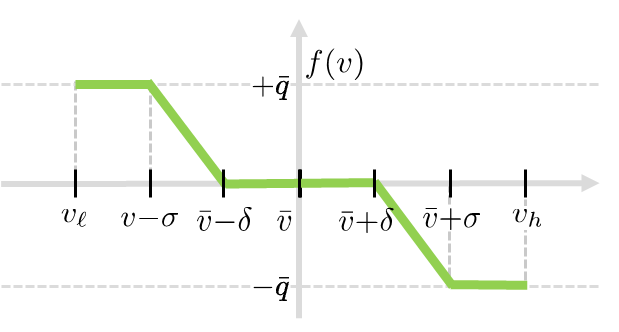}
    \vspace*{-1em}
	\caption{Volt/VAR control rule provisioned by the IEEE Standard 1547~\cite{IEEE1547}.}
	\label{fig:vvr}
\end{figure}

Upon reviewing some modeling preliminaries from~\cite{MGCK23}, this section formulates ORD. Per the Volt/VAR mode of the IEEE Std. 1547~\cite{IEEE1547}, each DER decides its reactive injection as a non-increasing and piecewise-linear function $q=f(v)$ of its local voltage like the one shown in Fig.~\ref{fig:vvr}. In general, the Volt/VAR rule can described by four breakpoints. Without harming generality, the Volt/VAR rule is assumed to be odd symmetric around voltage $\bar{v}$. Such symmetry `regularizes' ORD in case not many scenarios fall within a particular segment. The symmetric rule can be described by four parameters $(\bar{v},\delta,\sigma,\bar{q})$. The last three parameters relate to the negative slope $\alpha$ of non-flat segments as
\begin{equation}\label{eq:slope}
\alpha=\frac{\overline{q}}{\sigma-\delta}>0.
\end{equation}
The standard specifies the general shape of the rule by imposing the ensuing constraints on rule parameters~\cite{IEEE1547}:
\begin{subequations}\label{eq:1547con}
\begin{align}
0.95 &\leq \bar{v} \leq 1.05\label{eq:1547con:v}\\
0 &\leq   \delta     \leq 0.03\label{eq:1547con:delta}\\
\delta+0.02  &\leq   \sigma     \leq 0.18\label{eq:1547con:sigma}\\
0&\leq \overline{q}\leq \hat{q}.\label{eq:1547con:q}
\end{align}
\end{subequations}
Here $\hat{q}$ is the DER's reactive power capability. Because DERs should be able to offer Volt/VAR control regardless of their kW output, their kVA rating is usually oversized by 10\% their kW rating $\hat{p}$, so that $\hat{q}=\sqrt{1.1^2-1}\cdot \hat{p}=0.46\cdot \hat{p}$. 

Even though the standard determines the general shape of the rule, the exact values of $(\bar{v},\alpha,\delta,\sigma)$ are left for the operator to decide. Rule parameters can be customized per bus $n$, in which case, they are denoted by $(\bar{v}_n,\alpha_n,\delta_n,\sigma_n)$. Let vector $\bz$ carry all rule parameters across all buses as
\begin{equation}\label{eq:}
    \bz=[\bbv;~\balpha;~\bdelta;~\bsigma]
\end{equation}
where vectors $(\bbv,\balpha,\bdelta,\bsigma)$ collect parameters across buses. The goal is to customize $\bz$ via a ORD. Before doing so, let us briefly review the postulated grid model.

Consider a radial single-phase distribution system having $N+1$ buses indexed by $n=0,\ldots,N$. The substation bus is indexed by $n=0$ and the rest of the buses form set $\mcN:=\{1,\ldots,N\}$. Let vectors $\bv$ and $\tbp$ collect the voltage magnitudes and net active power injections at all buses in $\mcN$. The net reactive injections across $\mcN$ can be decomposed as $\bq+\tbq$, where vector $\bq$ denotes the controllable reactive injections by DERs, and $\tbq$ the uncontrollable ones. Given the vector of \emph{grid loading conditions} 
\begin{equation}\label{eq:tbtheta}
\btheta:=[\tbp;~\tbq]  
\end{equation}
the goal is to control $\bq$ via Volt/VAR rules to regulate voltage $\bv$ within allowable limits while minimizing ohmic losses.

Voltage magnitudes $\bv$ and total ohmic losses $\ell$ on lines are the two quantities of interest. We would like to express them as functions of $(\bq;\btheta)$. Lacking closed-form expressions and avoiding computationally demanding models, one can resort to Taylor's expansions around the point $\btheta=\bzero$ and $\bq=\bzero$. Under a first-order expansion, voltages can be expressed as~\cite{9091863}
\begin{equation}\label{eq:ldf}
\bv\simeq \bX\bq+\tbv(\btheta)~~\text{where}~~\tbv:=\bR\tbp+\bX\tbq+v_0\bone
\end{equation}
where $(\bR,\bX)$ are positive definite matrices depending on feeder topology, and $v_0$ is the substation voltage. Vector $\tbv$ denotes the voltages induced by $\btheta$ with no reactive injections by DERs. Under a second-order expansion, ohmic  losses can be approximated by the convex quadratic function~\cite{TJKT-SG21}
\begin{equation}\label{eq:losses}
\ell=(\bq+\tbq)^\top\bR(\bq+\tbq) + \tbp^\top\bR\tbp.
\end{equation}

When DERs controlled by Volt/VAR rules interact with the feeder, they give rise to nonlinear discrete-time dynamics:
\begin{subequations}\label{eq:dynamics}
\begin{align}
\bv^t &= \bX\bq^t + \tbv\label{eq:dynamics:v}\\
q_n^{t+1} &= f_n(v_n^t),\quad \forall n\in\mcN.\label{eq:dynamics:q}
\end{align}
\end{subequations}
The Volt/VAR dynamics of \eqref{eq:dynamics} are globally exponentially stable and feature a unique equilibrium provided $\|\diag(\balpha)\bX\|_2<1$; see e.g.,~\cite{9091863,Pedram13}. Diagonal matrix $\diag(\balpha)$ carries vector $\balpha$ on its main diagonal and $\|\cdot\|_2$ is the spectral norm. To avoid strict inequalities, the condition can be tightened as
$\|\diag(\balpha)\bX\|_2\leq 1-\epsilon$
for a small positive $\epsilon$. To lighten the computational burden, we have previously put forth an inner approximation of the latter constraint that is~\cite{MGCK23}
\begin{equation}\label{eq:stability2}
\bX\balpha \leq (1-\epsilon) \bone \quad \text{and} \quad 
\alpha_n\leq \frac{{1-\epsilon}}{\sum_{m\in\mcN}X_{nm}},~\forall n\in\mcN.
\end{equation}
which yields simpler linear inequality constraints. Stability conditions for multiphase feeders are provided in~\cite{GCK23}.


Stability condition~\eqref{eq:stability2} together with constraints \eqref{eq:1547con} outlining the shape of the rules form \emph{feasible set} $\mcZ$ for $\bz$:
\begin{equation}\label{eq:Z}
\mcZ:=\{\bz:\bz~\text{satisfying}~\eqref{eq:1547con},\eqref{eq:stability2}\}.
\end{equation}

Under~\eqref{eq:stability2}, Volt/VAR dynamics converge to \emph{equilibrium injections}, denoted by $\bq_{\bz}(\btheta)$ to emphasize their dependence on $\bz$ and $\btheta$. Equilibrium injections correspond to \emph{equilibrium voltages} $\bv_{\bz}(\btheta)=\bX\bq_{\bz}(\btheta)+\tbv(\btheta)$.

\subsection{Chance-Constrained Optimal Rule Design (ORD)}\label{subsec:ccord}
Control rules have been set up to avoid frequent communication with the operator. Suppose rules are designed every two hours. Given the uncertainty in $\btheta$, we postulate ORD as the chance-constrained program
\begin{align}\tag{\textsf{ORD}}\label{eq:ord}
\min_{\bz\in\mcZ}~&~\mathbb{E}\left[\ell(\bq_{\bz}(\btheta))\right]\\
\textrm{subject to}~&~\mathrm{Pr}\left(v_{\ell}\leq v_{n}\left(\bq_{\bz}(\btheta)\right)\leq v_{h}\right)\geq 1-\beta,~\forall n\in\mcN\nonumber
\end{align}
where $(v_{\ell},v_{h})$ are the allowable voltage limits. The probability of voltage limit violation $\beta$ can be set to a small percentage like $\beta=5\%$. The expectation and probability operators are taken with respect to uncertain grid conditions $\btheta$. Problem~\eqref{eq:ord} aspires to design Volt/VAR rules that are stable, comply with the IEEE Std. 1547, minimize the average ohmic losses, and maintain voltages within allowable limits for more than $(1-\beta)\%$ of the grid conditions. Parameter $\beta$ strikes a balance between ohmic losses and voltage excursions.

\begin{figure}[t]
	\centering
	\includegraphics[width=0.35\textwidth]{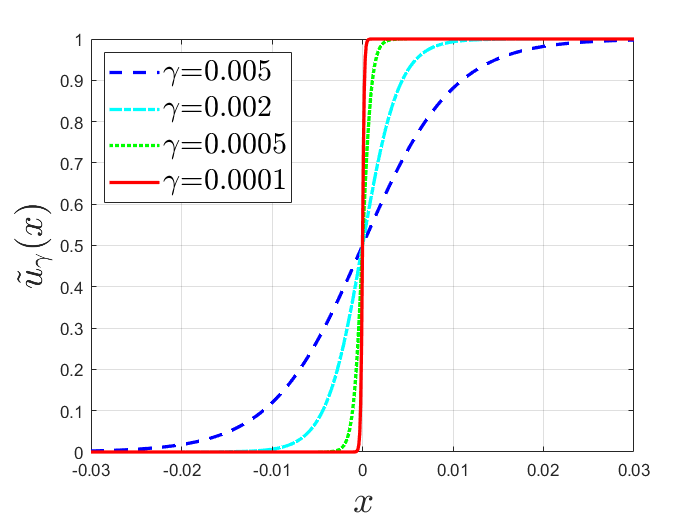}
	\vspace*{-1em}
    \caption{Logistic function approximation of the step function.}
	\label{fig:logisticVSstep}
\end{figure}

To cope with \eqref{eq:ord}, let us reformulate the chance constraints in two steps. First, convert double-sided voltage constraints to single-sided. If $v_{\ell}=0.97$ and $v_{h}=1.03$~pu, the voltage constraint can be expressed as
\[|v_{n}-1|\leq 0.03\quad \Leftrightarrow\quad \left(v_{n}-1\right)^2\leq 0.03^2\]
where $v_n$ here is the shorthand of $v_{n}\left(\bq_{\bz}(\btheta)\right)$. Squaring waives non-differentiability issues in later developments. Second, we replace the probability with an expectation operator. If $x$ is a random variable, it is easy to verify that $\mathrm{Pr}(x\geq 0)=\mathbb{E}\left[u(x)\right]$, where $u(x)$ is the \emph{unit step function}
\begin{equation}\label{eq:indicator}
u(x) = 
    \begin{cases}
    1, \quad x\geq 0 \\
    0, \quad x<0
    \end{cases}.  
\end{equation}  
Hence, the chance constraint in \eqref{eq:ord} can be expressed as
\begin{equation*}
\mathbb{E}\left[u\left(0.03^2-\left(v_{n}-1\right)^2\right)\right] \geq 1-\beta.
\end{equation*}
Since $1-u(x)=u(-x)$, the above is equivalent to
\begin{equation}\label{eq:cc1}
\mathbb{E}\left[u\left(\left(v_{n}-1\right)^2-0.03^2\right)\right] \leq \beta, \quad \forall n \in \mcN.
\end{equation}
Because $u(x)$ is discontinuous at $x=0$ and has zero gradients elsewhere, it is not a convenient option for gradient-based solvers. To overcome this limitation, we surrogate the step by the (mirrored) logistic function shown in Fig.~\ref{fig:logisticVSstep}. The logistic function and its derivative are
\begin{equation*}
    \tilde{u}_{\gamma}(x)\coloneqq \frac{1}{1+e^{-x/\gamma}} \quad \text{and}\quad 
    \frac{\d{\tilde{u}_{\gamma}}}{\d x}= \frac{1}{\gamma} \tilde{u}_{\gamma}(x)\left(1-\tilde{u}_{\gamma}(x)\right)
\end{equation*} 
with constant $\gamma>0$ controlling the approximation accuracy. Function $\tilde{u}_{\gamma}(x)$ approaches $u(x)$ as $\gamma\rightarrow 0^+$. However, for smaller values of $\gamma$, the logistic approximation inherits the derivative complications of $u(x)$ as its derivative approaches infinity. Setting $\gamma=10^{-4}$ seemed to strike a good trade-off between accuracy and convergence during our numerical tests. Using the logistic function, constraint \eqref{eq:cc1} can be replaced by 
\begin{align}\label{eq:cc2}
\mathbb{E}\left[\tilde{u}_{\gamma}\left((v_{n}-1)^2-0.03^2\right)\right] \leq \beta, \quad \forall n \in \mcN.
\end{align}
We next present how \eqref{eq:ord} can be handled leveraging tools from deep learning and stochastic approximation.


\section{Proposed Methodology}\label{sec:method}
Even after surrogating chance constraints, problem~\eqref{eq:ord} faces two challenges to be addressed in this section: 
\renewcommand{\theenumi}{\emph{c\arabic{enumi}}}
\begin{enumerate}
    \item It considers voltages and losses at \emph{steady state}, i.e., after Volt/VAR dynamics of \eqref{eq:dynamics} have settled; and
    \item The distribution of $\btheta$ is unknown. Even if known, evaluating cost and constraint functions would not be easy.
\end{enumerate}

\subsection{An RNN Emulator of Volt/VAR Dynamics}\label{subsec:emulator}

\begin{figure}[t]
	\centering
	\includegraphics[scale=0.3]{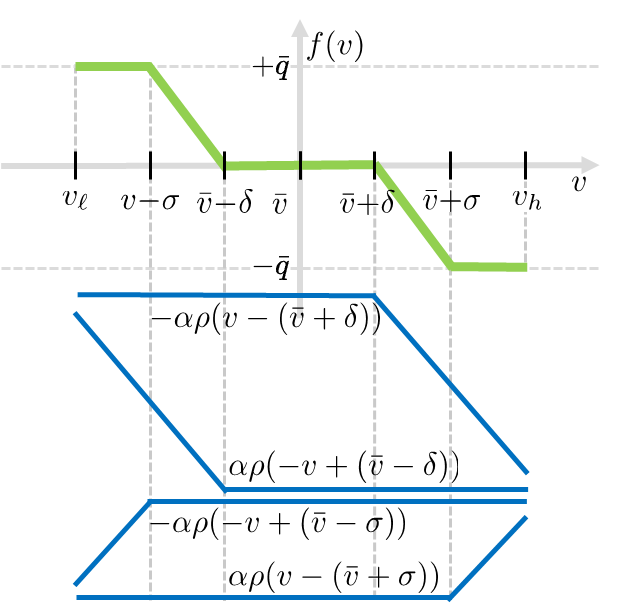}
	\caption{Volt/VAR rule $f(v)$ as a linear combination of four ReLUs.}
	\label{fig:relus}
\end{figure}

\begin{figure}[t]
\centering
	\includegraphics[scale=0.5]{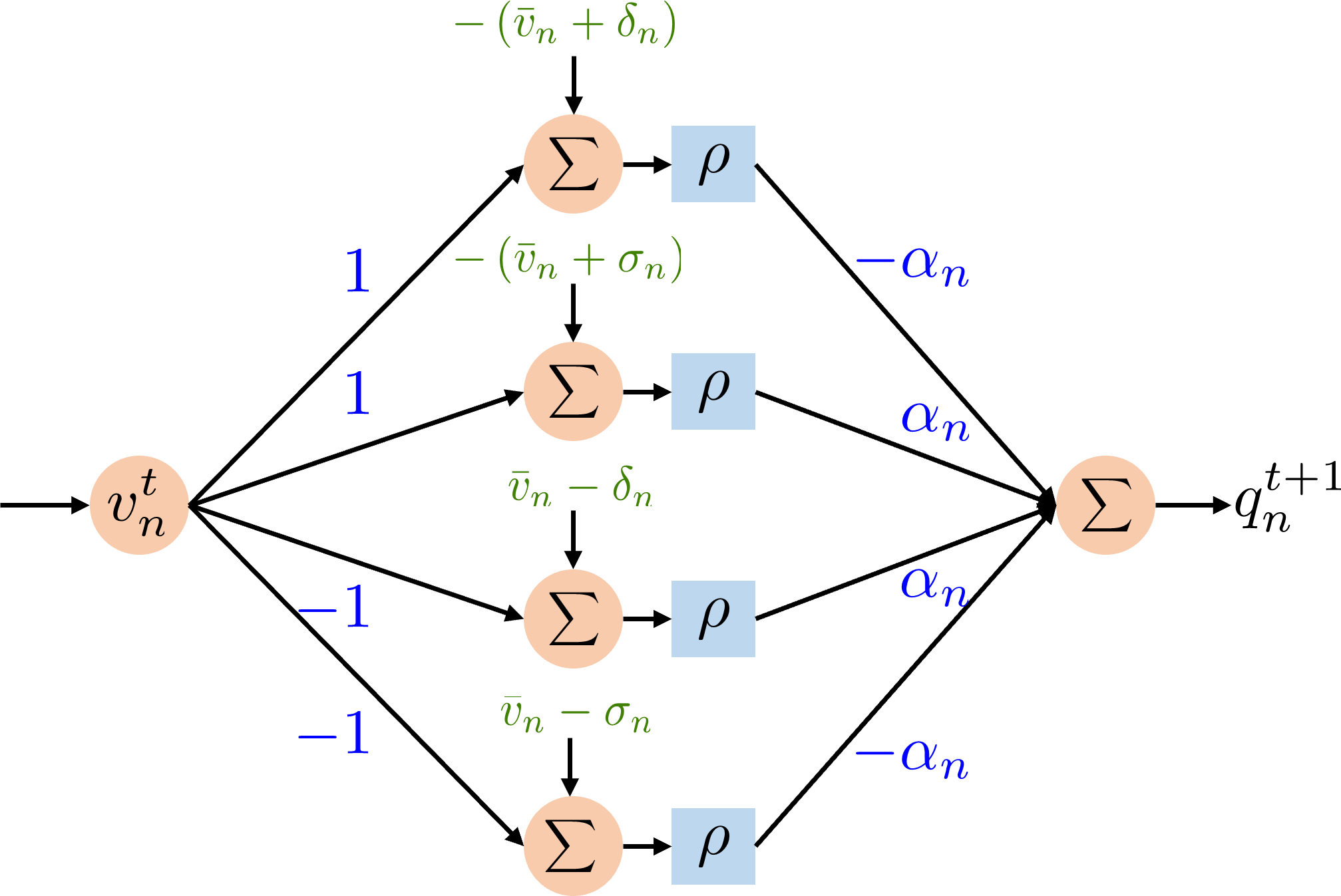}
	\caption{Volt/VAR rule $f_n(v_n)$ for DER $n$ as a single hidden-layer NN.}
	\label{fig:single_unit}
\end{figure}

Regarding \emph{c1)}, to describe equilibrium injections $\bq_{\bz}(\btheta)$, one has to use a mixed-integer nonlinear program (MINLP). We have previously developed and experimented with such MINLP for an unconstrained version of ORD. The MINLP scaled unfavorably with the problem size, and was outperformed by a DNN-based approach. The MINLP can only be further challenged in the setup considered in this work, due to chance constraints. To this end, we adopt the DNN-based approach of~\cite{GCK23}, which is briefly reviewed for completeness.

\begin{figure}[t]
\centering
	\includegraphics[scale=0.2]{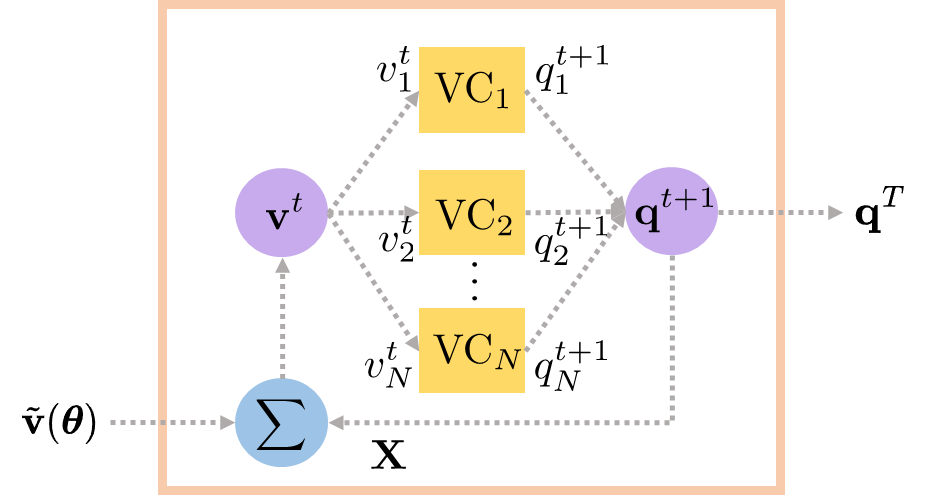}
	\caption{The RNN emulating Volt/VAR dynamics. The yellow blocks termed $\text{VC}_n$ implement one pass of the Volt/VAR rule of Fig.~\ref{fig:vvr} for each DER $n$ using the single-layer NN of Fig.~\ref{fig:relus}. The RNN input is the vector of uncompensated voltages $\tbv(\btheta)=\bR\tbp+\bX\tbq+v_0\bone$, which depends on scenario $\btheta=[\tbp;~\tbq]$. RNN weights depend linearly on vector $\bz$ of $4N$ rule parameters. The RNN output is the reactive injection obtained at step $T$ of Volt/VAR dynamics. For large enough $T$, the RNN output approximates equilibrium $\bq_{\bz}(\btheta)$.}
	\label{fig:rnn}
\end{figure}

The main idea is that the dynamics of \eqref{eq:dynamics} can be emulated by a \emph{recursive neural network} (RNN). A key observation is that the rule of Fig.~\ref{fig:vvr} can be expressed as a linear combination of unit ramp functions $\rho(x)=x\cdot u(x)$ or \emph{rectified linear units (ReLUs)}, depicted in Fig.~\ref{fig:relus}. Thanks to this decomposition, rule $f_n(v_n)$ of DER $n$ can be emulated by the single hidden-layer NN of Fig.~\ref{fig:single_unit}. This NN takes $v_n^t$ as input, parameters $(\bar{v}_n,\delta_n,\sigma_n,\alpha_n)$ as weights, and computes reactive injection $q_n^{t+1}$. Denote this NN as $\text{VC}_n$ for all $n$. Figure~\ref{fig:rnn} shows how $\text{VC}_n$'s can be combined to yield Volt/VAR dynamics: They can be stacked vertically so that given $\bv^t$, they compute $\bq^{t+1}$ per \eqref{eq:dynamics:q}. Reactive injections are subsequently passed through an affine layer of fixed weights to emulate $\bv^{t+1}=\bX\bq^{t+1}+\tbv$ as in \eqref{eq:dynamics:v}. This structure is repeated $T$ times (depth $2T$) to mimic $T$ steps of Volt/VAR dynamics. 


The RNN takes vector $\tbv$ of uncompensated voltages as input, which depends on grid scenario $\btheta$ per \eqref{eq:ldf}. At its output, the RNN computes reactive injections $\bq^T$ at time step $T$. A simple adaptation of~\cite[Prop.~1]{GCK23} shows that if
\[T\geq \frac{\log 2\|\hbq\|_2- \log \epsilon_1}{\log(1-\epsilon)^{-1}}\]
where $\hbq$ has kVAR ratings, the RNN output $\bq^T_{\bz}(\btheta)$ is $\epsilon_1$ close to $\bq_{\bz}(\btheta)$ in the sense $\|\bq^T-\bq_{\bz}(\btheta)\|_2\leq \epsilon_1$. The RNN has only $4N$ tunable weights as they repeat at every layer. The advantage of using a RNN as a Volt/VAR emulator is that given $(\bz,\btheta)$, one can swiftly compute equilibrium injections and their gradients with respect to $\bz$ due to gradient back-propagation. This computational efficiency stems from the proficiency of modern deep learning libraries, such as PyTorch. We next put forth an algorithm for training the RNN of Fig.~\ref{fig:rnn} so its weights $\bz$ solve \eqref{eq:saord} approximately.

\subsection{RNN Training via a Primal-Dual Algorithm}\label{subsec:pd}
Challenge \emph{c2)} is addressed via stochastic sample approximation that replaces ensemble with sample averages as
\begin{equation*}
\mathbb{E}\left[\ell(\bq_{\bz}(\btheta))\right]\simeq \frac{1}{S}\sum_{s=1}^S\ell(\bq_{\bz}(\btheta_s))
\end{equation*} 
and likewise for the left-hand side of \eqref{eq:cc2}. Suppose the operator has a set of $S$ scenarios $\{\btheta_s\}_{s=1}^S$, which are representative of the grid conditions anticipated for the next two hours. Using sample approximations, problem \eqref{eq:ord} is reformulated as
\begin{align}\tag{\textsf{SA-ORD}}\label{eq:saord}
\min_{\bz\in\mcZ}~&~\frac{1}{S}\sum_{s=1}^S\ell(\bq_{\bz}(\btheta_s))\\
\textrm{subject to}~&~
\frac{1}{S}\sum_{s=1}^Sg_n(\bz;\btheta_s)\leq \beta,~\forall n\in\mcN.\nonumber
\end{align}
Here we have used the shorthand notation:
\begin{equation*}
g_n(\bz;\btheta_s):=\tilde{u}_{\gamma}\left((v_{n}(\bq_{\bz}(\btheta_s))-1)^2-0.03^2\right)~\forall n\in\mcN.
\end{equation*}
Problem \eqref{eq:saord} is solved via a primal-dual decomposition algorithm as explained next. Let vector $\blambda$ collect the Lagrange multipliers $\lambda_n$ associated with the $n$-th voltage constraint. The Lagrangian function of \eqref{eq:saord} is
\begin{equation*}
L(\bz;\blambda)=\frac{1}{S}\sum_{s=1}^S\ell(\bq_{\bz}(\btheta_s))+\sum_{n\in\mcN}\lambda_n\left(\frac{1}{S}\sum_{s=1}^S g_n(\bz;\btheta_s)-\beta\right)
\end{equation*}
Primal-dual decomposition aims at minimizing $L(\bz;\blambda)$ over $\bz\in\mcZ$ and maximizing it over $\blambda\geq \bzero$. The algorithm involves iterations indexed by $k$ with each iteration having two updates.

\emph{Primal update:} This is a projected gradient descent step on $L(\bz;\blambda^k)$ with respect to $\bz$. It updates the primal variables as
\begin{equation}\label{eq:pd:p}
\bz^{k+1}=\left[\bz^k-\mu_z\nabla_{\bz}L(\bz^k;\blambda^k)\right]_{\mcZ}    
\end{equation}
where $[\cdot]_{\mcZ}$ denotes projection onto $\mcZ$, and $\mu_z>0$ is the step size. Due to~\eqref{eq:1547con:q}, the feasible set $\mcZ$ is not convex. Nonetheless, constraints \eqref{eq:1547con} and \eqref{eq:stability2} can be shown to be convex with respect to variable $\cbz=[\bbv;~\bc;~\bdelta;~\bsigma]$, where $\bc$ has entries $c_n=1/\alpha_n$ for all $n$; see~\cite{MGCK23}. Since the $\bz-\cbz$ mapping is a one-to-one, the update in \eqref{eq:pd:p} can be implemented in four steps:
\renewcommand{\theenumi}{\emph{s\arabic{enumi}}}
\begin{enumerate}
    \item The RNN weight vector $\bz^k$ is updated to the intermediate vector $\hbz^{k}:=\bz^k-\mu_z\nabla_{\bz}L(\bz^k;\blambda^k)$. This step is accomplished with the aid of RNN libraries. 
    \item Vector $\hbz^{k}$ be mapped to an alternative space where each $\alpha_n$ has been converted to $c_n=1/\alpha_n$. Symbol $\cbz^k$ denotes the transformed vector.
    \item Project $\cbz^k$ onto constraints \eqref{eq:1547con} and \eqref{eq:stability2}. This can be posed as a second-order cone program (SOCP)~\cite{MGCK23}. Let the obtained vector be $\cbz^{k+1}$. 
    \item Vector $\cbz^{k+1}$ is transformed back to the original space by inverting its $c_n$ entries to $\alpha_n=1/c_n$ for all $n$. The obtained vector is $\bz^{k+1}$ appearing in \eqref{eq:pd:p}.
\end{enumerate}

\emph{Dual update:} The second step of iteration $k$ is a projected gradient ascent on $L(\bz^{k+1};\blambda)$ with respect to $\blambda$. It updates the dual variables (Lagrange multipliers) of chance constraints as:
\begin{equation}\label{eq:pd:d}
\lambda_n^{k+1}=\left[\lambda_n^k+\mu_{\lambda}\left(\frac{1}{S}\sum_{s=1}^S g_n(\bz^{k+1};\btheta_s)-\beta\right)\right]_+  
\end{equation}
where $[x]_+:=\max\{x,0\}$ and $\mu_{\lambda}>0$ is the step size. The dual update entails computing constraints. Equilibrium voltages can be computed from $\bq_{\bz^{k+1}}(\btheta_s)$ for all $s$. 


\section{Numerical Tests}\label{sec:tests}
\begin{figure}[t]
	\centering
	\includegraphics[scale=0.5]{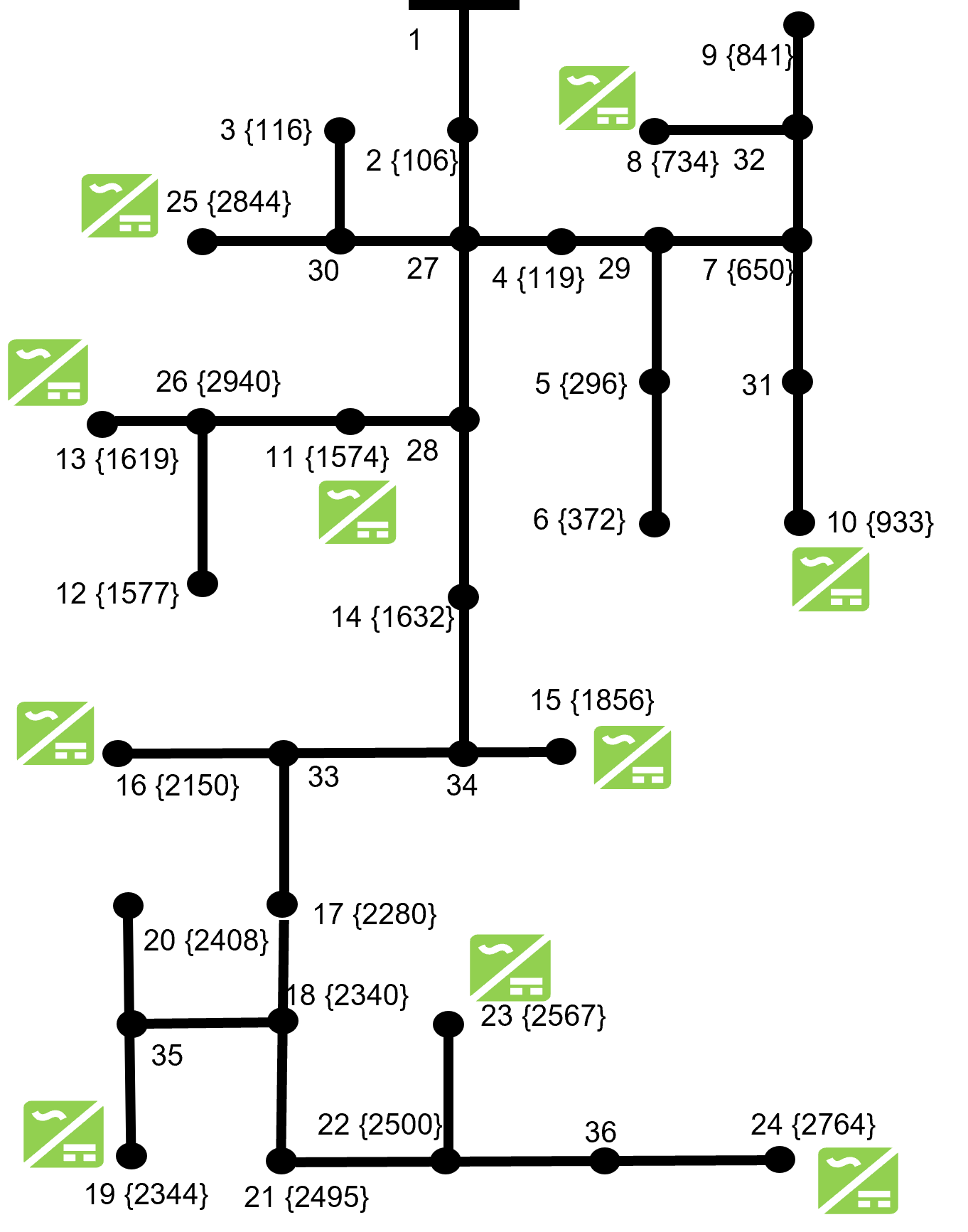}
	\caption{IEEE 37-bus feeder with 10 DERs.}
	\label{fig:37bus}
\end{figure}

The proposed ORD was tested on the single-phase equivalent of the IEEE 37-bus feeder. Real-world data of active load and solar generation at one-minute resolution were downloaded from the Smart* project for the day of April 2, 2011~\cite{Smartsolar}. The dataset consists of active loads from 443 homes and generation from $43$ solar panels. Loads from multiple homes were averaged to better approximate loads at primary network nodes. Homes with IDs $20$-$369$ were averaged 10 at a time, and successively added as active loads to buses $2$-$26$ as shown in Fig.~\ref{fig:37bus}. Each averaged load was normalized so its peak value during the day coincided with the nominal active power load of its hosting node. For each time interval, reactive loads were added by randomly sampling lagging power factors within $[0.9,1]$. Each solar generation signal was normalized so its peak value was 1.6 times that of the nominal active load of the hosting bus. Active generation from solar panels was also added, as per the mapping in Fig.~\ref{fig:37bus}. Apparent power limits for inverters were set to $1.1$ times the peak active generation.

Control rules were designed and evaluated in Python 3.10 on a 2.9~Ghz base speed AMD Ryzen 8-core processor laptop computer with 16~GB RAM. Pytorch was selected as the library to design and train RNNs, as it implements computation graphs dynamically, in which the number of layers $T$ does not need to be fixed beforehand but is decided on the fly. The criterion for not adding an additional layer to the RNN was met as long as the $\ell_2$-norm between two consecutive layers became smaller than $10^{-7}$. All RNNs were trained using the Adam optimizer. The projection step \emph{s3)} was solved as a SOCP using the CVXPY library in Python with Gurobi as the solver. Rule parameters were initialized as $(\bar{v}_n,\delta_n,\sigma_n,\alpha_n)=(1,0.01,0.03,1.5)$ for all $n$. We sampled $S=80$ scenarios from the high-solar period 3:00 -- 4:20 PM. Step sizes were selected as $\mu_z= 0.001\cdot0.99^{k}$ and $\mu_{\lambda}= 0.0015\cdot0.99^{k}$. The stability margin was set as $\epsilon=0.5$.

\begin{table}[t]
    \centering
    \caption{Worst-Case Voltage Violation Probabilities and Ohmic Losses}
    \begin{tabular}{|c|c|c|}
    \hline\hline
     & Worst-Case  &  Ohmic Losses \\
     &  Chance Probability [\%] &  [$\times 10^{-2}$] \\
    \hline\hline
       no VAR by DERs & $62.5$ & $2.89$ \\
    \hline
       default Volt/VAR rules & $55.0$  & $2.95$ \\
    \hline
       ORD with $\beta=0.20$ & $20.0$  & $3.26$ \\
    \hline
       ORD with $\beta=0.15$ & $15.0$  & $3.37$ \\
    \hline
       ORD with $\beta=0.10$ & $11.25$  & $3.48$ \\
    \hline
       ORD with $\beta=0.05$  & $5.0$   & $3.61$ \\
    \hline\hline
    \end{tabular}
    \label{tab:beta}
\end{table}

\begin{figure}[t]
	\centering
	\includegraphics[scale=0.4]{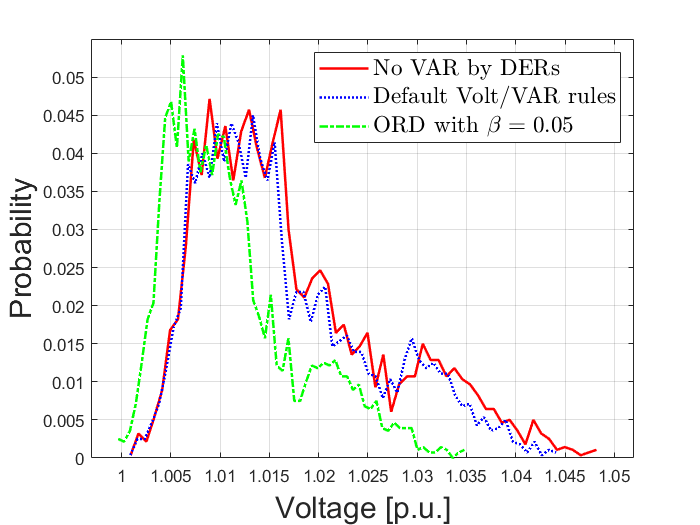}
 	\vspace*{-0.5em}
	\caption{Histograms of voltages across all buses and scenarios.}
	\label{fig:beta=0.05}
\end{figure}

The performance of our ORD was evaluated in terms of ohmic losses and voltage chance probabilities as shown in Table~\ref{tab:beta}. Chance probabilities are reported in terms of their maximum value over buses. With no reactive injections from DERs, the voltage at one of the buses lies outside the range $[0.97,1.03]$ for $62.5\%$ of the scenarios. The chance probability reduces to $55\%$ for the worst-case bus by using the default values for $\bz$ recommended by the IEEE Std. 1547, namely $\bar{v}_{n}=1$, $\delta_{n}=0.02$, $\sigma_{n}=0.08$, $\bar{q}_{n}=\hat{q}_{n}$. Via ORD, the worst-case chance probability reduces with decreasing $\beta$. Evidently, the algorithm was able to achieve near-optimal rule designs. Ohmic losses increase as voltage profiles improve. This is because the feeder experiences overvoltage issues so DERs are requested through the rules to consume rather than inject kVAR. To examine the voltage profile across the feeder, Fig.~\ref{fig:beta=0.05} depicts the empirical probability distribution of voltages across all buses and scenarios. ORD manages to shift the voltage profile closer to unity. 

\begin{figure}[t]
	\centering
	\includegraphics[scale=0.4]{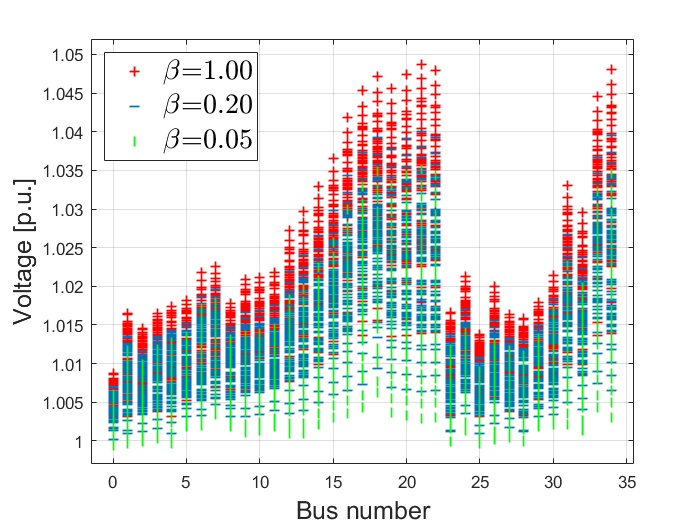}
 	\vspace*{-0.5em}
	\caption{Voltage profiles for $S=80$ scenarios for different values of $\beta$.}
	\label{fig:volt3}
\end{figure}

\begin{table}[t]
    \centering
    \caption{Errors in Voltages Between Linearized and Exact AC Models}
    \vspace*{-0.5em}
    \begin{tabular}{|c|c|c|}
    \hline\hline
         & Mean Error & Maximum Error  \\
    \hline\hline
       $\beta=0.20$  & $7.94\cdot10^{-4}$ & $2.74\cdot10^{-3}$ \\
    \hline
       $\beta=0.15$  & $7.93\cdot10^{-4}$  & $2.78\cdot10^{-3}$\\
    \hline
       $\beta=0.10$  & $7.88\cdot10^{-4}$  & $2.73\cdot10^{-3}$\\
    \hline
       $\beta=0.05$  & $8.12\cdot10^{-4}$  & $2.76\cdot10^{-3}$  \\
    \hline\hline
    \end{tabular}
    \label{tab:DCAC}
    \vspace*{-1.5em}
\end{table}

To explore the effect of $\beta$, Fig.~\ref{fig:volt3} shows how voltages improve for decreasing $\beta$. For $\beta=1.00$, ORD ignores voltage constraints and aims for minimizing losses. The designed $\bz$ was also evaluated on the exact AC grid model. Upon initializing at $\bq=\bzero$, equilibrium injections and voltages were computed for all scenarios. Table~\ref{tab:DCAC} reports the errors between equilibrium voltages computed by the two models, and showcases the linearized model is sufficiently accurate. 


\section{Conclusions}\label{sec:conclusions}
A novel chance-constrained ORD formulation to minimize ohmic losses subject to voltage constraints has been put forth. Uncertainty on grid conditions has been coped with via sample approximations and chance constraints. The piecewise-affine nature of the rules has been handled through a RNN emulator of Volt/VAR dynamics, whose weights are optimized through primal-dual decomposition to find near-optimal rule parameters. Tests using real-world data on a benchmark feeder confirm the proposed method can address ORD over a range of loading conditions, satisfy voltage chance constraints, and that the linearized grid model is not far off the AC model. Open research directions include generalizations to multiphase feeders, incorporating AC feeder models into the RNN, as well as exploring incremental and other rules.

\balance
\bibliographystyle{IEEEtran}
\bibliography{myabrv,kekatos,inverters}

\begin{thebibliography}{10}
\providecommand{\url}[1]{#1}
\csname url@samestyle\endcsname
\providecommand{\newblock}{\relax}
\providecommand{\bibinfo}[2]{#2}
\providecommand{\BIBentrySTDinterwordspacing}{\spaceskip=0pt\relax}
\providecommand{\BIBentryALTinterwordstretchfactor}{4}
\providecommand{\BIBentryALTinterwordspacing}{\spaceskip=\fontdimen2\font plus
\BIBentryALTinterwordstretchfactor\fontdimen3\font minus
  \fontdimen4\font\relax}
\providecommand{\BIBforeignlanguage}[2]{{%
\expandafter\ifx\csname l@#1\endcsname\relax
\typeout{** WARNING: IEEEtran.bst: No hyphenation pattern has been}%
\typeout{** loaded for the language `#1'. Using the pattern for}%
\typeout{** the default language instead.}%
\else
\language=\csname l@#1\endcsname
\fi
#2}}
\providecommand{\BIBdecl}{\relax}
\BIBdecl

\bibitem{IEEE1547}
\emph{IEEE Standard for Interconnection and Interoperability of DERs with
  Associated Electric Power Systems Interfaces}, IEEE Std., 2018.

\bibitem{9091863}
X.~Zhou, M.~Farivar, Z.~Liu, L.~Chen, and S.~H. Low, ``Reverse and forward
  engineering of local voltage control in distribution networks,'' \emph{{IEEE}
  Trans. Autom. Contr.}, vol.~66, no.~3, pp. 1116--1128, 2021.

\bibitem{Jabr18}
R.~A. Jabr, ``Linear decision rules for control of reactive power by
  distributed photovoltaic generators,'' \emph{{IEEE} Trans. Power Syst.},
  vol.~33, no.~2, pp. 2165--2174, Mar. 2019.

\bibitem{SKL19GM}
M.~K. Singh, V.~Kekatos, and C.-C. Liu, ``Optimal distribution system
  restoration with microgrids and distributed generators,'' in \emph{Proc.
  {IEEE} Power \& Energy Society General Meeting}, Atlanta, GA, Aug. 2019.

\bibitem{GCK23}
S.~Gupta, S.~Chatzivasileiadis, and V.~Kekatos, ``Deep learning for optimal
  {Volt/VAR} control using distributed energy resources,'' \emph{{IEEE} Trans.
  Smart Grid}, 2023, https://arxiv.org/abs/2210.12805.

\bibitem{Pedram13}
P.~Jahangiri and D.~C. Aliprantis, ``Distributed {Volt/VAr} control by {PV}
  inverters,'' \emph{{IEEE} Trans. Power Syst.}, vol.~28, no.~3, pp.
  3429--3439, Aug. 2013.

\bibitem{9796576}
A.~Savasci, A.~Inaolaji, and S.~Paudyal, ``Two-stage {Volt-VAR} optimization of
  distribution grids with smart inverters and legacy devices,'' \emph{{IEEE}
  Trans. Ind. Applicat.}, vol.~58, no.~5, pp. 5711--5723, 2022.

\bibitem{PaudyalMISOCP}
------, ``Distribution grid optimal power flow with adaptive {Volt-VAr} droop
  of smart inverters,'' in \emph{IEEE Industry Applications Society Annual
  Meeting}, Vancouver, BC, 2021, pp. 1--8.

\bibitem{Paudyal-LinDist3Flow}
A.~Inaolaji, A.~Savasci, and S.~Paudyal, ``Distribution grid optimal power flow
  in unbalanced multiphase networks with {Volt-VAr} and {Volt-Watt} droop
  settings of smart inverters,'' \emph{{IEEE} Trans. Ind. Applicat.}, vol.~58,
  no.~5, pp. 5832--5843, 2022.

\bibitem{8365842}
A.~Singhal, V.~Ajjarapu, J.~Fuller, and J.~Hansen, ``Real-time local {Volt/Var}
  control under external disturbances with high {PV} penetration,''
  \emph{{IEEE} Trans. Smart Grid}, vol.~10, no.~4, pp. 3849--3859, Jul. 2019.

\bibitem{9781808}
J.~Sepulveda, A.~Angulo, F.~Mancilla-David, and A.~Street, ``Robust
  co-optimization of droop and affine policy parameters in active distribution
  systems with high penetration of photovoltaic generation,'' \emph{{IEEE}
  Trans. Smart Grid}, vol.~13, no.~6, pp. 4355--4366, 2022.

\bibitem{Baker18}
K.~Baker, A.~Bernstein, E.~Dall'Anese, and C.~Zhao, ``Network-cognizant voltage
  droop control for distribution grids,'' \emph{{IEEE} Trans. Power Syst.},
  vol.~33, no.~2, pp. 2098--2108, Mar. 2018.

\bibitem{9609090}
R.~Xu, C.~Zhang, Y.~Xu, Z.~Dong, and R.~Zhang, ``Multi-objective
  hierarchically-coordinated volt/var control for active distribution networks
  with droop-controlled pv inverters,'' \emph{{IEEE} Trans. Smart Grid},
  vol.~13, no.~2, pp. 998--1011, Mar 2022.

\bibitem{MGCK23}
I.~Murzakhanov, S.~Gupta, S.~Chatzivasileiadis, and V.~Kekatos, ``Optimal
  design of {Volt/VAR} control rules for inverter-interfaced distributed energy
  resources,'' \emph{{IEEE} Trans. Smart Grid}, 2023, (early access).

\bibitem{Baosen22}
W.~Cui, J.~Li, and B.~Zhang, ``Decentralized safe reinforcement learning for
  inverter-based voltage control,'' \emph{Electric Power Systems Research},
  vol. 211, p. 108609, 2022.

\bibitem{10106014}
Z.~Yuan, G.~Cavraro, M.~K. Singh, and J.~Cortes, ``Learning provably stable
  local {Volt/Var} controllers for efficient network operation,'' \emph{{IEEE}
  Trans. Power Syst.}, pp. 1--14, 2023.

\bibitem{Turitsyn11}
K.~Turitsyn, P.~Sulc, S.~Backhaus, and M.~Chertkov, ``Options for control of
  reactive power by distributed photovoltaic generators,'' \emph{Proc. {IEEE}},
  vol.~99, no.~6, pp. 1063--1073, Jun. 2011.

\bibitem{Ayyagari17}
K.~S. Ayyagari, N.~Gatsis, and A.~F. Taha, ``Chance-constrained optimization of
  distributed energy resources via affine policies,'' in \emph{{IEEE
  GlobalSIP}}, Montreal, Canada, Nov. 2017, pp. 1050--1054.

\bibitem{Li18}
P.~{Li}, B.~{Jin}, D.~{Wang}, and B.~{Zhang}, ``Distribution system voltage
  control under uncertainties using tractable chance constraints,''
  \emph{{IEEE} Trans. Power Syst.}, vol.~34, no.~6, pp. 5208--5216, Nov. 2019.

\bibitem{TJKT-SG21}
S.~Taheri, M.~Jalali, V.~Kekatos, and L.~Tong, ``Fast probabilistic hosting
  capacity analysis for active distribution systems,'' \emph{{IEEE} Trans.
  Smart Grid}, vol.~12, no.~3, pp. 2000--2012, May 2021.

\bibitem{Smartsolar}
D.~Chen, S.~Iyengar, D.~Irwin, and P.~Shenoy, ``Sunspot: Exposing the location
  of anonymous solar-powered homes,'' in \emph{ACM Intl. Conf. on Systems for
  Energy-Efficient Built Environ.}, Palo Alto, CA, Nov. 2016.

\end{thebibliography}
\end{document}